\documentclass[runningheads]{llncs}
\usepackage[T1]{fontenc}
\usepackage{braket}
\usepackage{bigints}
\usepackage{graphicx}
\begin{document}
\sloppy
%
\title{Sonification of Wigner functions: case study of intense light-matter interactions}
\titlerunning{Sonification of Wigner functions}
%

\author{Reiko Yamada\inst{1} \and
Antoine Reserbat-Plantey\inst{3}\orcidID{0000-0002-9106-8750} \and
Eloy Piñol\inst{1}\orcidID{0000-0002-3763-175X} \and
Maciej Lewenstein\inst{1,2}\orcidID{0000-0002-0210-7800}}

\authorrunning{R. Yamada et al.}
%

\institute{ICFO -- Institute of Photonic Science, Barcelona Institute of Science and Technology Av. C.F. Gauss, 3 - 08860 - Castelldefels - Barcelona - Spain \and  ICREA Lluis Companys, 23 - 08010 - Barcelona - Spain \and Université Côte d’Azur, CNRS, CRHEA. Valbonne, Sophia-Antipolis, France}

\maketitle              
\begin{abstract}
In quantum mechanics, the Wigner function $\rho_W(\textbf{r},\textbf{p})$ serves as a phase-space representation, capturing information about both the position $\textbf{r}$ and momentum $\textbf{p}$ of a quantum system. 
The Wigner function facilitates the calculation of expectation values of observables, examination of quantum system dynamics, and analysis of coherence and correlations. 
Therefore, it might serve as a tool to express quantum systems intuitively, for example, by using sonification techniques.
This paper summarizes the experimental strategies employed in a previous project and delineates a new approach based on its outcomes. 
Emphasizing the attribution of specific Wigner functions to their underlying quantum states, dynamics, and sources; our proposed methodology seeks to refine the sonification and scoring process, aiming to enhance intuitive understanding and interpretation of quantum phenomena.
\end{abstract}
The development of atomic and molecular optics\cite{srakaew_subwavelength_2023}, ultra-fast optics\cite{lewenstein_generation_2021}, nanophotonics\cite{hensen_loophole-free_2015} or quantum transport – to name a few – has led to experimental situations involving genuine and complex quantum mechanics phenomena. 
While quantum mechanics textbooks often refer to the building blocks such as quantum harmonic oscillator, single particle in a box, or single spin states, modern physics has to deal with systems with physical properties that evolve on a very short length scale, often smaller than the De Broglie wavelength\cite{cohen-tannoudji_quantum_1997}.
Consequently, quantum phenomena such as entanglement, cluster states, mesoscopic quantum state superposition, or strongly correlated states appear. 
Their description and modeling sometimes require mathematical tools, such as Wigner functions\cite{cohen-tannoudji_quantum_2020}.
Sonification of quantum phenomena has often revolved around simple physical systems, such as a Gaussian wavepacket in a potential well\cite{cadiz_sound_2014}, or quantum harmonic oscillators\cite{hutchison_quantum_2010,duplessis_rodneydupqhosyn_2023}.  
Motivations for sonifying quantum phenomena vary, from enhancing understanding of quantum behavior \cite{cadiz_sound_2014} to developing sound-based protocols for quantum information storage\cite{hutchison_quantum_2010} or even creating tools for sorting and categorizing features within quantum systems\cite{campo_sonification_2005}.
Campo's 2005 study \cite{campo_sonification_2005} elegantly demonstrated the sonification of mass spectra of subatomic particles like baryons.
Some approaches also emphasize the aesthetic aspect of sonifying quantum phenomena, exemplified by Rodney Du Plessis' piece \textit{Psi} (2021), which utilizes the \textit{QHOSYN} sound emulator to simulate quantum harmonic oscillators' superposition \cite{duplessis_rodneydupqhosyn_2023}. 
In such a manner, this tool enables the simulation and generation of quantum-based sound.
Another illustration comes from the sonification of many-body systems – which are situations often challenging to analyze –, as presented in Bob Sturm's piece \textit{50 Particles} (2001), where musical gestures map inter-particles coupling, and the system dynamics drive the piece development\cite{sturm_composing_2001}. 

As this article will detail, Wigner functions are powerful tools for describing and computing massively quantum systems. 
We attempted to sonify the Wigner function describing a quantum superposition of optical states – known as Schr\"odinger cat state – using various methodologies, and we anticipate that sonifying its quantum nature may yield unexpected manifestations beyond preconceived paradigms.
 Therefore, our objective is to explore various strategies for sonifying quantum phenomena, to achieve a multi-perspective understanding, exemplified by optical Schr\"odinger cat states\cite{lewenstein_generation_2021} through auditory information. 

Simultaneously, we are interested in providing an auditory perspective on the Wigner function that faithfully describes the system as it exists rather than concentrating solely on specific studies of it.
In some respects, our approach echoes the sentiment expressed by John Cage\cite{cage_silence_2012}: “I have nothing to say, and I am saying it, and that is poetry as I need it”. 

We will begin by examining Wigner functions' utilization to depict highly non-classical states. 
Subsequently, we will describe the methodology for obtaining datasets within a specific experimental context: intense light-matter interactions aimed at generating Schr\"odinger cat states. 
Finally, we will introduce various sonification strategies and scoring processes employed in our approach.

\section{Wigner distributions}
\subsection{Classical distributions and quasi-distributions}
In classical mechanics, a particle's position $r$ and momentum $p$ can be determined with arbitrary precision. 
If the particle's state is defined statistically, it can be described by a distribution function, $\rho_{cl}(r,p)$. 
This function is always positive, must be normalized to unity, and may include correlations between the particle's position and velocity. 

In quantum mechanics, these two representations – position and momentum space (connected through Fourier transformation) – are also commonly used. 
In contrast to the classical situation, these representations are exclusive. 
All information regarding the particle's momentum is lost in the position representation, while the converse is also true.
A necessary consequence is that no information about correlations between the particle's position and momentum can be extracted. 

The latest quantum mechanical description is often not practical, and there is a need for an intermediate description that maintains information about $r$ and $p$ – with a specific limitation in the precision – while respecting the laws of quantum mechanics.
Such a tool was introduced by Wigner in 1932 and consists of a quantum mechanical function – named $\rho_W(r,p)$ – that allows computing average values, similar to what can be done classically. 
Indeed, in a classical distribution $\rho_{cl}(r,p)$, if one wants to know the average value of a given operator $\mathcal{A}$, the distribution shall be multiplied by this operator and summed over all phase space $(r,p)$.
Wigner distributions – or Wigner functions – are thus helpful in computing any average values of observables. 
From a practical standpoint, the Wigner function contains all information about a quantum system and its evolution. 
This article considers sonifying Wigner functions to gain a more intuitive perception of a quantum system and its dynamics.
Therefore, we shall first describe the key differences and similarities between a classical statistical distribution function and a Wigner quasi-distribution function. 

\subsection{Negative values in Wigner functions}
While classical distributions are always positive or zero and normalized to 1, the description of a quantum system can be more nuanced. 
On the one hand, there are cases where the Wigner function $\rho_W(r,p)$ is always positive and thus behaves like a classical distribution.
Here, such Wigner function can be seen as a probabilistic distribution and is used to describe systems having quantities that vary in space over large distances (i.e. on a length scale much larger than the de Broglie wavelength $\lambda = h/\|\textbf{p}\|$), as found in quantum electronic transport.
Notably, to fulfill the Heisenberg uncertainty principle, the width of the Wigner function around a given value – i.e. $p_0(r)$ – is of the order of $\hbar/d$, where $d$ is the coherence length of the wavefunction.
In systems where physical quantities vary at a length scale comparable to $\lambda$, quantum effects become important.
Surprisingly, the Wigner distribution can take negative values. 
As a result, it is no longer possible to refer to it as a probability distribution, and the more appropriate term is quasi-distribution. 
It is often said that negative values of the Wigner function carry the quantumness of a physical system. 
We will derive the Wigner function for a case study and explain the emergence and signification of negative values. 
We consider the first excited state $\Psi_1(t)$ of a 1D quantum harmonic oscillator.
The envelope of this odd wavefunction is similar to the first harmonics of a string, having a node at its center. 
Its Wigner function at $(r,p)=(0,0)$ is defined by\cite{cohen-tannoudji_quantum_2020}: 
\begin{eqnarray}
\rho_W (r=0,p=0)&=\frac{1}{2\pi\hbar}\bigintss dy \ \Psi_1\left(\frac{y}{2}\right)\Psi_1^*\left(-\frac{y}{2}\right) \nonumber \\
&=-\frac{1}{2\pi\hbar}\bigintss dy \ |\Psi_1\left(\frac{y}{2}\right)|^2<0
\end{eqnarray}
The negative value highlights that a Wigner function cannot be seen as a probability distribution but rather a quasi-distribution. 
Negative values for $\rho_W$ are often observed in various situations, ranging from simple interference experiments like Young's slits (tamasic or dark rays) to complex quantum states resulting from intense laser-atom interactions\cite{lewenstein_generation_2021}. 

\section{Data from intense light-matter interactions experiments}
\begin{figure}[ht]
\centering
\includegraphics[width=\textwidth]{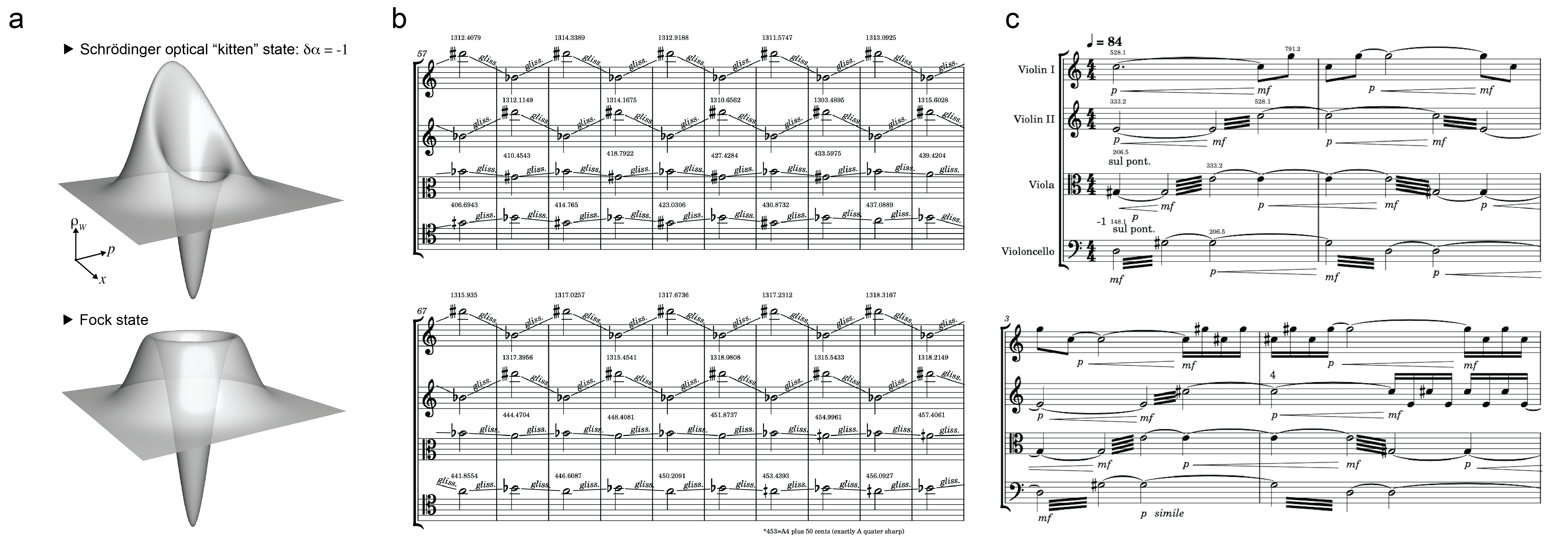}
\caption{\textbf{Sonification of Wigner functions}. \textbf{a}: Phase space $\left(r,p\right)$ representation of the Wigner functions $\rho_W\left(r,p\right)$ of a Schr\"odinger kitten state for $\delta\alpha=-1$ (top, Eq.\ref{eq:cat}), and the first Fock state (bottom, Eq. \ref{eq:fock}) are plotted. \textbf{b}: Excerpt of the score for string quartet using the method II. \textbf{c}: Excerpt of the score for string quartet using the method III. Both methods II and III are described in Section \ref{sec:method}.} 
\label{fig1}
\end{figure}

Our data is obtained through the collaboration between theoreticians and experimentalists who study exotic quantum states that originate from the interaction of intense laser fields with atomic clouds\cite{lewenstein_generation_2021}.
Especially for our second sonification approach described in the methodology section, a comprehensive understanding of the experimental conditions \cite{lewenstein_generation_2021,stammer_high_2022} is imperative.
A coherent laser state initially interacts with a cloud of $N$ Xenon atoms in their ground state. 
This interaction produces higher harmonic states and amplitude-shifted coherent states.
The quantum superposition of both states can produce an optical Schr\"odinger cat state – in this case, kitten state –, similar to the quantum superposition of at least two different macroscopic or mesoscopic states, such as $\ket{cat \ alive}$ and $\ket{cat \ dead}$. 
Their Wigner function has the form:
\begin{eqnarray}
\rho_W^{Cat}(\beta)&=&\frac{2}{\pi(1-e^{(-|\delta\alpha|^2)})}[e^{(-2|\beta-\alpha-\delta\alpha|^2 )} \nonumber \\ 
&+&e^{(-|\delta\alpha|^2 )} e^{(-2|\beta-\alpha|^2 )} 
-e^{(-|\delta\alpha|^2)}e^{(-2|\beta-\alpha|^2)} \nonumber \\
&&(e^{(2(\beta-\alpha)\delta\alpha^* )}+e^{(2(\beta-\alpha)^*\delta\alpha)})].
\label{eq:cat}
\end{eqnarray}
Here, $\alpha$ and $\beta$ are complex numbers related to position and momentum through the expression: $\beta -\alpha = r+ip$. 
The value $\delta \alpha$ reflects the shift in amplitude for the fundamental state. 
This shift is due to correlations between the fundamental and the generated harmonics and is central to understanding system evolution. 
When this shift is large, e.g. $\delta \alpha<-3$, the resulting state is a coherent state with a Gaussian profile, almost analog to a purely classical state.
However, when $\delta \alpha$ goes from $-1$ and tends to $0$, the state evolves from a Schr\"odinger cat state (Eq. \ref{eq:cat}) to a quantum state with a well-defined number of photons, also known as Fock state. 
The Wigner function describing such Fock state is:
\begin{eqnarray}
\rho_W^{F_m} (r,p)&=\frac{(-1)^m}{\pi} e^{-(r^2+p^2 )} {\cal L}_m \left(2(r^2+p^2)\right)\nonumber\\
\rho_W^{F_1} (r,p) &=-\frac{1}{\pi} e^{-(r^2+p^2 )} \left(1 - 2(r^2+p^2)\right)
\label{eq:fock}
\end{eqnarray}
where ${\cal L}_m$ denotes the $m$th Laguerre polynomial. 
Both Wigner functions for the Schr\"odinger cat state (Eq. \ref{eq:cat}) and the first Fock state (Eq. \ref{eq:fock}) are shown in Fig. \ref{fig1}a.

\section{Sonification approaches to Wigner functions}
\subsection{Mesh discretization method of Wigner quasi-distribution}
In our previous trials\cite{yamada_towards_2023}, we explored direct sound generation from Wigner functions, employing a mapping method that discretizes its arguments and values into grids, ensuring 99 \% coverage of the distribution. 
This mesh discretization strongly depends on the type of grid used. 
Therefore, we considered two approaches: regular grids with all the lattice sites equidistant and non-regular grids based on Gaussian intervals, as detailed previously\cite{yamada_towards_2023}. 
The latter phase-space discretization accurately pictures the variations of the Wigner function as it reduces the binning size to gain sampling resolution. 

\subsection{Sound generation}\label{sec:method}
These meshes were subsequently translated into auditory representations through various techniques: (I) mapping grid points to phase, frequency, and amplitude of waves, generating up to 900 concurrent sinusoidal waves; (II) directly mapping minimum and maximum Wigner function values to frequencies; (III) segmenting Wigner function values in four sections and correlating with frequency intensity; and (IV) mapping mathematical moments of the Wigner function to a set of harmonic oscillators distributed over a Gaussian envelope. 

The neutral sinusoidal waves and triangular pulses controlled by Wigner function data were used for acousmatic rendering. 
For acoustic instruments, frequencies were rounded to the nearest quarter division of 12-note equal-tempered pitches, and different playing techniques were employed to control timbres, reflecting the negativity of the Wigner function (shown in Fig. \ref{fig1}b-c).

In the method IV, we compute the central value $r_0$ and the standard deviation $\sigma_r$ of $\rho_W(r,p)$. 
These parameters were then associated with the central frequency $f_0$ and the spectral filter bandwidth (often denoted as Q-value in EQ treatment) of the generated sound event.
Each event was produced from a Gaussian distribution centered at $f_0(r_0)$ of $N$ different harmonic oscillators.
The envelope of this Gaussian distribution has a bandwidth mapped by $Q(\sigma_r)$.
The resulting sonogram (or Fourier space representation of the generated sound) is displayed in Fig. \ref{fig2}a.
Here, one can distinguish a Gaussian envelope for a set of $N=21$ oscillators (appearing as a bright set of lines at low frequencies) from upper harmonics. 
The central value of the Gaussian distribution increases linearly as the Wigner function second moment $\sigma_r(\delta \alpha)$ changes, according to our mapping protocol. Notably, the individual oscillator bandwidth remains constant throughout this process.

The resulting audio events in every case exhibited a high degree of dissonance, characterized by transformations in the nuances of dissonance and a rich, sustained texture, generating a plethora of overtones. 
This auditory experience starkly contrasts with the prevalent visual representation of the Wigner function. 
Depending on the mapping variables employed, massively quantum states (generally understood as when $\rho_W$ displays negative values) can be distinguished from classical states through sound, albeit discernible only to those informed about the mapping approach to these variables.

\subsection{Integrating Wigner function's evolution into electro-acoustic medium}
\begin{figure}[ht]
\centering
\includegraphics[width=\textwidth]{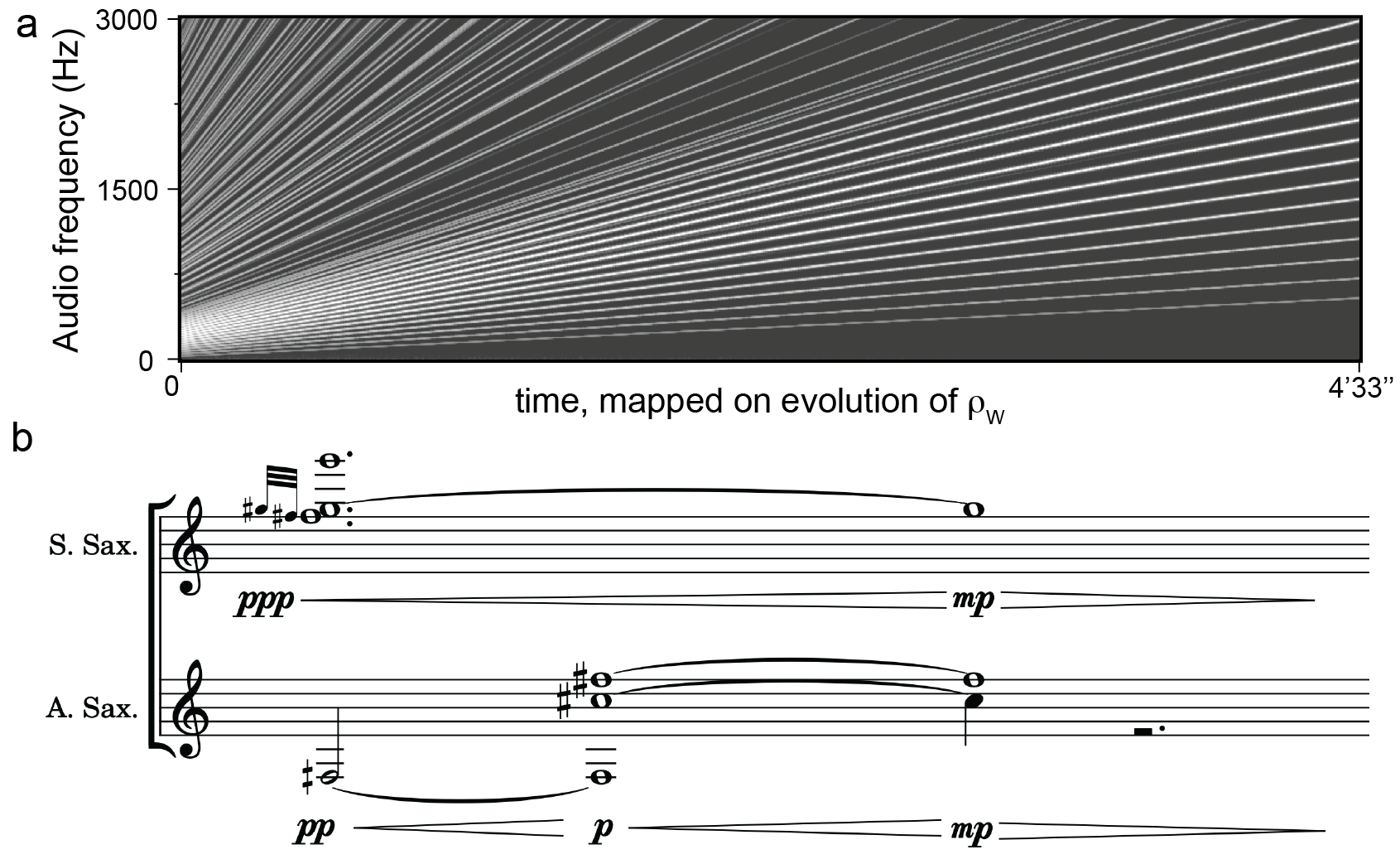}
\caption{\textbf{Spectral analysis of the rendition IV}. \textbf{a}: Audio sonogram representation (reduced to monophonic) of the transition from the first Fock state to a kitten state and then to a coherent state in 4'33". \textbf{b}: Excerpt from the score for soprano and alto saxophones using multiphonics.} 
\label{fig2}
\end{figure}

In our latest methodology, we augment the diverse expressions of the Wigner function with a focused consideration of its experimental conditions. 
We integrate the method IV, as described in the previous section, using single-reed wind instrument multiphonics to articulate the superposition of fundamental and high harmonics in our experiments. 
Specifically, in this latest method, we combine the functional elements derived from the quasi-distribution $\rho_W$ – namely, the first ($r_0$) and second ($\sigma_r$) moments –, with the multiphonics of wind instruments (Fig. \ref{fig2}b). 
This integration gives a dual perception of $\rho_W$ and its corresponding experimental situation.
Indeed, on the one hand, one can track the evolution of the Wigner function $\rho_W$ as a function of $\delta \alpha$, transiting from a Fock state ($\delta\alpha \rightarrow 0$) to a kitten state ($\delta\alpha=-1$), and finally to an almost classical coherent state ($\delta\alpha=-3$). 
On the other hand, multiphonic sound generation in single-reed wind instruments maps the experimental process of quantum superposition of higher harmonic states and amplitude-shifted coherent laser states.
Thus, this integration enables us to simultaneously and audibly track the system's evolution. 
Furthermore, the position and momentum information encoded in the Wigner function $\rho_W(r,p)$ can be directly utilized for multi-channel spatialization in live performance settings. 

\section{Conclusion and outlook}
Highly non-classical states, such as the ones studied here and originating from intense light-matter interactions, require mathematical tools such as Wigner functions for description and computation.
While attributing physical significance to the Wigner function must be cautious, they prove invaluable for computing average values of quantum observables and representing the physical properties of a given system \cite{cohen-tannoudji_quantum_2020}. 
In this work, we introduced various strategies to sonify Wigner functions to offer an auditory perspective. 
In particular, we proposed a mesh discretization method for accurately mapping Wigner functions and sonifying $\rho_W$ from four perspectives based on direct parameter mapping (I-III). 
Furthermore, in some cases, the scoring (e.g. the use of \textit{sul ponticello} or ricochet when $\rho_W<0$) highlighted the negative values of the Wigner function.

The latest sonifying method IV, based on extracting the moments of $\rho_W$, offers a statistical analysis of the Wigner quasi-distribution. 
Our study was limited to computing the two first moments (mean value and standard deviation). 
Additionally, extracting the third and fourth moments – skewness and kurtosis, respectively – could provide further insights into the asymmetric behavior (in $(r,p)$ space) and the tail extension of the Wigner quasi-distribution $\rho_W$. 

One exciting area of research involves tackling interference phenomena, such as the Young double-slit experiment. 
In this experiment, the wavefunction of a particle (such as a photon) is split into two coherent wavepackets situated at $r=-a$ and $r=a$ after passing through the slits. 
The diagonal components of the corresponding Wigner function describe the two wavepackets. 
In contrast, the off-diagonal components are non-zero at $r=0$, where the probability of finding a particle is zero. 
These off-diagonal components, also known as a \textit{ghost}\cite{cohen-tannoudji_quantum_2020} components, explain the observed interference pattern when the wavepackets overlap. 
This scenario presents opportunities for sonification, as the system's observable can be precisely calculated using Wigner functions.
\section*{Acknowledgments}
\scriptsize The academic and artistic research mentioned in this article was made possible by ICFO and Phonos Foundation (Barcelona, Spain). The authors thank Philipp Stammer (ICFO) for his generous help with topics related to the Wigner function. R.Y., E.P. and M.L. acknowledge support from: ERC AdG NOQIA; MCIN/AEI (PGC2018-0910.13039/501100011033,  CEX2019-000910-S 10.13039/501100011033, Plan National FIDEUA PID2019-106901GB-I00, Plan National STAMEENA PID2022-139099NB-I00 project funded by MCIN/AEI/10.13039 501100011033 and by the “European Union NextGenerationEU/PRTR” (PRTR-C17.I1), FPI); QUANTERA MAQS PCI2019 111828-2); QUANTERA DYNAMITE PCI2022 132919 (QuantERA II Programme co-funded by European Union’s Horizon 2020 program under Grant Agreement No 101017733),  Ministry for Digital Transformation and of Civil Service of the Spanish Government through the QUANTUM ENIA project call - Quantum Spain project, and by the European Union through the Recovery, Transformation and Resilience Plan - NextGenerationEU within the framework of the Digital Spain 2026 Agenda; Fundació Cellex; Fundació Mir-Puig; Generalitat de Catalunya (European Social Fund FEDER and CERCA program, AGAUR Grant No. 2021 SGR 01452, QuantumCAT/U16-011424, co-funded by ERDF Operational Program of Catalonia 2014-2020); Barcelona Supercomputing Center MareNostrum (FI-2023-1-0013); EU Quan-tum Flagship (PASQuanS2.1, 101113690) EU Horizon 2020 FET-OPEN OPTOlogic (Grant No 899794); EU Horizon Europe Program (Grant Agreement 101080086 - NeQST), results incorporated in this standard have received funding from the European Innovation Coun-cil and SMEs Executive Agency under the European Union’s Hori-zon Europe programme), ICFO Internal “QuantumGaudi” project; European Union’s Horizon 2020 program under the Marie Sklodow-ska-Curie grant agreement No 847648;  La Caixa Junior Leaders fellowships, La Caixa Foundation (ID 100010434): CF/BQ/PR23/11980043. Views and opinions expressed are, however, those of the author(s) only and do not necessarily reflect those of the European Union, European Commission, European Climate, Infra-structure and Environment Executive Agency (CINEA), or any other granting authority. Neither the European Union nor any granting authority can be held responsible for them. E.P. has been supported by Ayuda (PRE2021-098926) funded by MCIN/AEI/ 10.13039/501100011033 and by FSE+. 
\bibliographystyle{splncs04}
\bibliography{Wigner_sonification.bib}
\end{document}